\documentclass[%
reprint,
superscriptaddress,
nofootinbib,
]{revtex4-1}
\usepackage{units}
\usepackage{graphicx}
\usepackage{gensymb}

\newcommand{\msub}[1]{_{\text{#1}}}
\newcommand{\musb}[1]{_{\text{#1}}}

\newcommand{\avg}[1]{\left< #1 \right>} 
\newcommand{\refF}[1]{Fig.\,\ref{#1}}
\newcommand{\refT}[1]{Tab.\,\ref{#1}}
\newcommand{\refE}[1]{Eq.\,\ref{#1}}

\usepackage[usenames, dvipsnames]{xcolor}

\begin{document}

\title{CRITICAL FIELDS OF Nb$_3$Sn PREPARED FOR SUPERCONDUCTING CAVITIES}

\author{S. Keckert} \email{Sebastian.Keckert@helmholtz-berlin.de} \affiliation{Helmholtz-Zentrum Berlin f\"ur Materialien und Energie GmbH (HZB), Berlin, Germany} \affiliation{Universit\"at Siegen, Siegen, Germany}

\author{T. Junginger} \email{Tobias.Junginger@lancaster.ac.uk} \email{now at Lancaster University and the Cockcroft Institute, Warrington, UK} \affiliation{Helmholtz-Zentrum Berlin f\"ur Materialien und Energie GmbH (HZB), Berlin, Germany} \affiliation{TRIUMF Canada's National Laboratory for Particle and Nuclear Physics, Vancouver, Canada}

\author{T. Buck}	\affiliation{TRIUMF Canada's National Laboratory for Particle and Nuclear Physics, Vancouver, Canada} 
\author{D. Hall}	\affiliation{Cornell Laboratory for Accelerator-based Sciences and Education, Ithaca, USA} 

\author{P. Kolb}	\affiliation{TRIUMF Canada's National Laboratory for Particle and Nuclear Physics, Vancouver, Canada} 
\author{O. Kugeler}	\affiliation{Helmholtz-Zentrum Berlin f\"ur Materialien und Energie GmbH (HZB), Berlin, Germany}
\author{R. Laxdal}	\affiliation{TRIUMF Canada's National Laboratory for Particle and Nuclear Physics, Vancouver, Canada} 
\author{M. Liepe}	\affiliation{Cornell Laboratory for Accelerator-based Sciences and Education, Ithaca, USA}
\author{S. Posen}	\email{now at Fermi National Accelerator Laboratory, Batavia, USA} \affiliation{Cornell Laboratory for Accelerator-based Sciences and Education, Ithaca, USA}  
\author{T. Prokscha}	\affiliation{Paul Scherrer Institut (PSI), Villigen, Switzerland} 
\author{Z. Salman}	\affiliation{Paul Scherrer Institut (PSI), Villigen, Switzerland}
\author{A. Suter}	\affiliation{Paul Scherrer Institut (PSI), Villigen, Switzerland}

\author{J. Knobloch}	\affiliation{Helmholtz-Zentrum Berlin f\"ur Materialien und Energie GmbH (HZB), Berlin, Germany} \affiliation{Universit\"at Siegen, Siegen, Germany} 

\vspace{10pt}

\begin{abstract}
Nb$_3$Sn is currently the most promising material other than niobium for future superconducting radiofrequency cavities. Critical fields above \unit[120]{mT} in pulsed operation and about \unit[80]{mT} in CW have been achieved in cavity tests. This is large compared to the lower critical field as derived from the London penetration depth, extracted from low field surface impedance measurements. In this paper direct measurements of the London penetration depth from which the lower critical field and the superheating field are derived are presented. The field of first vortex penetration is measured under DC and RF fields. The combined results confirm that Nb$_3$Sn cavities are indeed operated in a metastable state above the lower critical field but are currently limited to a critical field well below the superheating field.
\end{abstract}


\maketitle 

\section{Introduction}

The performance of Nb as a material for superconducting radiofrequency (SRF) cavities is reaching fundamental limits in terms of surface resistance and peak surface magnetic field. Other superconductors are being developed to offer a way to push performance beyond Nb technology. One potential alternative material is Nb$_3$Sn, a Type-II superconductor with a relatively high Ginzburg Landau parameter $\kappa$ ($\approx40$ close to stoichiometry \cite{Godeke2006}) as opposed to about 1.4 for high purity Nb. Nb$_3$Sn has a relatively high critical temperature of about \unit[18]{K}, twice that of Nb which allows, for example, to operate RF cavities at \unit[4.2]{K} with the same losses as for Nb at \unit[2]{K}, if dominated by intrinsic losses from thermally activated quasiparticles, so called BCS losses. This is of high relevance, since cavities can be operated with liquid helium at atmospheric pressure which considerably simplifies the required cryo plant and reduces its operating and investment costs significantly. The lower critical field $\mu_0H\msub{c1}$ for Nb$_3$Sn is about \unit[38]{mT} close to stoichiometry, as opposed to \unit[170]{mT} for clean Nb. To obtain accelerating gradients above $\approx$\unit[10]{MV/m} for $\beta$=$1$ elliptical cavities Nb$_3$Sn has therefore either to be operated in the mixed phase, with the penetrated vortices being pinned to avoid strong dissipation, or in a metastable vortex free state. This state can potentially be sustained up to the superheating field $H\msub{sh}$ which is about \unit[440]{mT} for Nb$_3$Sn, compared to about \unit[240]{mT} for Nb. Superconducting material parameters at T=\unit[0]{K} relevant to this work of Nb and Nb$_3$Sn are presented in \refT{tab:MaterialParameters}. The values of $\kappa$ and $H\msub{c1}$ are for clean Nb and Nb$_3$Sn close to stoichiometry, while $H\msub{sh}$ is computed from \cite{PhysRevB.83.094505}
\begin{equation}
H\msub{sh}=H\msub{c}\left(\frac{\sqrt{20}}{6}+\frac{0.55}{\sqrt{\kappa}}\right).
\label{eq:Hsh}
\end{equation}

\begin{table}
	\centering
	\caption{Material parameters of Nb and Nb$_3$Sn.}
		\begin{tabular}{l|c|c}
		Property & Nb & Nb$_3$Sn \\
		\hline
		$T\msub{c}$ [K] & 9.25 \cite{finnemore1966superconducting} & 18 \cite{Godeke2006} \\
		$\kappa$(0K) & 1.4 \cite{finnemore1966superconducting} & 34 \cite{Godeke2006} \\
		$\xi_0$ [nm]  & 39 \cite{MattisBardeenTheory} & 5.7(0.6) \cite{orlando1979critical} \\
		$\lambda\msub{L}$ [nm] & 27(3) \cite{Suter2005} &  65\cite{Poole}-89\cite{Hein1999} \\
		$\mu_0 H\msub{c1}$(0K) [mT] & 174 \cite{finnemore1966superconducting} & 38 \cite{Godeke2006} \\
		$\mu_0 H\msub{c}$(0K) [mT] & 199 \cite{finnemore1966superconducting} & 520 \cite{Godeke2006} \\
		$\mu_0 H\msub{sh}$(0K) [mT] & 240 \cite{PhysRevB.83.094505} & 440 \cite{PhysRevB.83.094505} \\
		\end{tabular}
	\label{tab:MaterialParameters}
\end{table}

\begin{figure}[ht]
	\centering
	\includegraphics[width=0.95\columnwidth]{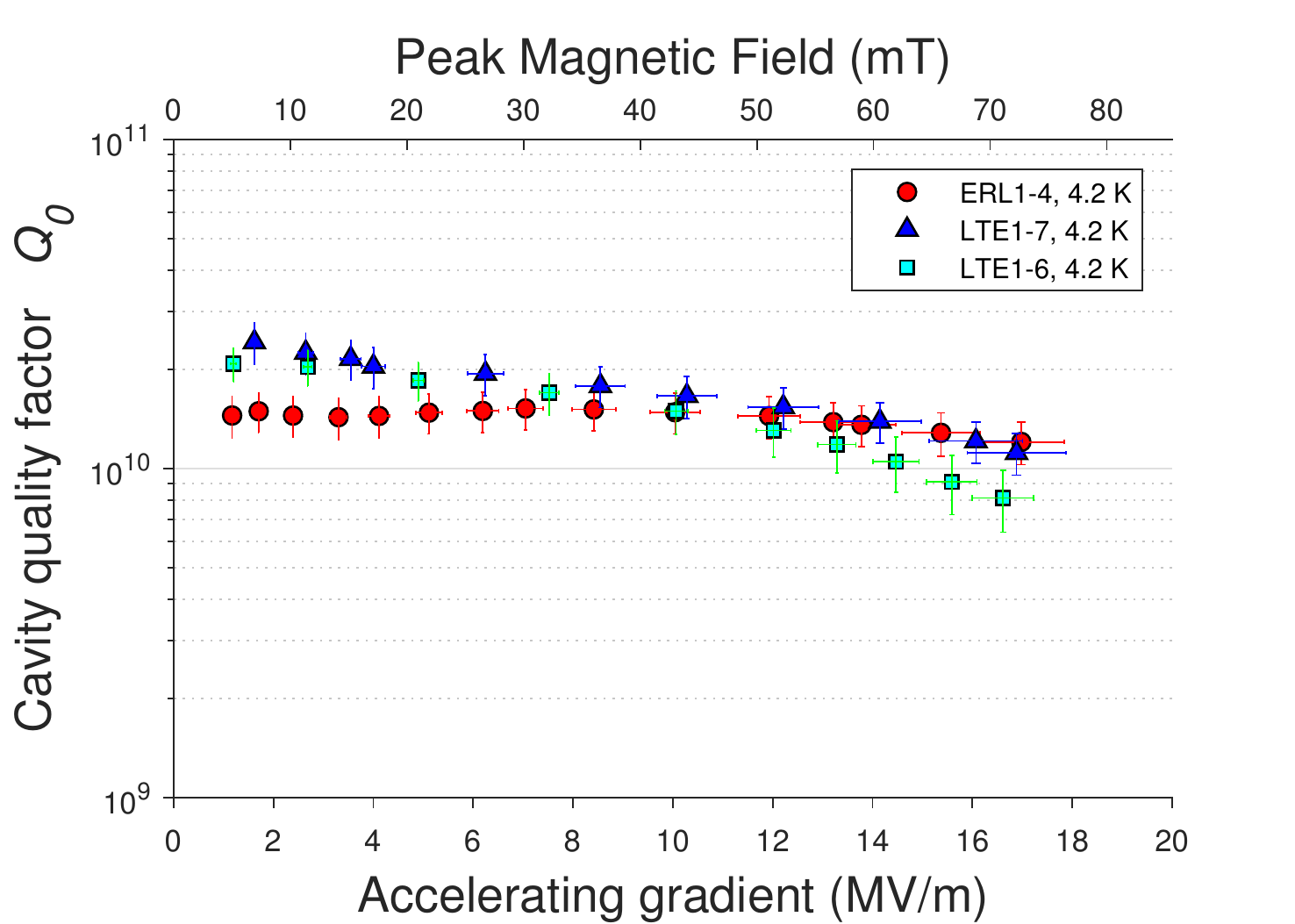}
	\caption{Cavity quality factor of three different Nb$_3$Sn coatings on \unit[1.3]{GHz} single-cell cavities, measured at \unit[4.2]{K}. Plot adapted from \cite{hall2017ipac}.}
	\label{fig:qvse}
\end{figure}

A Nb$_3$Sn SRF program began at Cornell University in 2009 \cite{posen2014prstab, posen2015apl}.
The program focused on coating fine grain bulk niobium cavities with a layer of $\unit[2]{\mu m}$ of Nb$_3$Sn via the deposition of tin vapor in a vacuum furnace. Cold tests of the cavities revealed (see \refF{fig:qvse}):
(1) reproducible high $Q_0$ values on the order of $10^{10}$ at \unit[4.2]{K}, (2) reproducible sustaining of this high $Q_0$ to useful accelerating gradients above \unit[16]{MV/m} corresponding to peak magnetic values in excess of \unit[70]{mT}. The attained $B\msub{p}$ values are significantly higher than $\mu_0H\msub{c1}$ as derived from the electron mean free path extracted from low field surface impedance measurements, suggesting that vortex penetration at $H\msub{c1}$ is not a limitation \cite{posen2015prl}. The scope of this paper is to test by direct measurements whether this conclusion is indeed substantiated. 

Using the low energy muon spin rotation (LE-$\mu$SR) technique from PSI the penetration depth $\lambda$ is measured directly. This allows to calculate $H\msub{c1}$. So called surface muons\footnote{These muons are called surface muons because they are produced from pions which stop and decay close to the surface of the production target.} available at TRIUMF are used to derive the DC field of first flux entry. The measured values are compared to the calculated $H\msub{c1}$. Finally, the critical RF field is tested using a Quadrupole Resonator (QPR) at HZB. Samples of specific shape adapted to the experiments have been produced using the same vapor diffusion technique as for the SRF cavities at Cornell.

\section{Sample preparation}
All samples are made from fine grain niobium of high residual resistivity ratio (RRR\textgreater300) and were coated with a $\unit[2]{\mu m}$ thick layer of Nb$_3$Sn at Cornell. The coating procedure for the used samples is the same as successfully used on the RF cavities \cite{posen2017nb3sn}. Coating lasts \unit[3]{hours}, with the cavity at \unit[1100]{$\degree$C} and the tin source at \unit[1200]{$\degree$C}.

\subsection{$\mu$SR samples}
For the LE-$\mu$SR experiments a coin-shaped sample was cut from a Nb sheet, while for the surface $\mu$SR measurements an ellipsoid sample is used which has been machined from bulk Nb. The coin is \unit[3]{mm} thick and \unit[20]{mm} in diameter. A hole is drilled through the center of the sample for mounting the sample during coating. The prolate ellipsoid is \unit[9]{mm} in radius on the minor axis with a \unit[22.9]{mm} semi-major axis. A threaded hole is drilled in one end for mounting the sample during coating and test. The muons are implanted at the opposite pole.

\begin{figure}[ht]
	\centering
		\includegraphics[width=0.90\columnwidth]{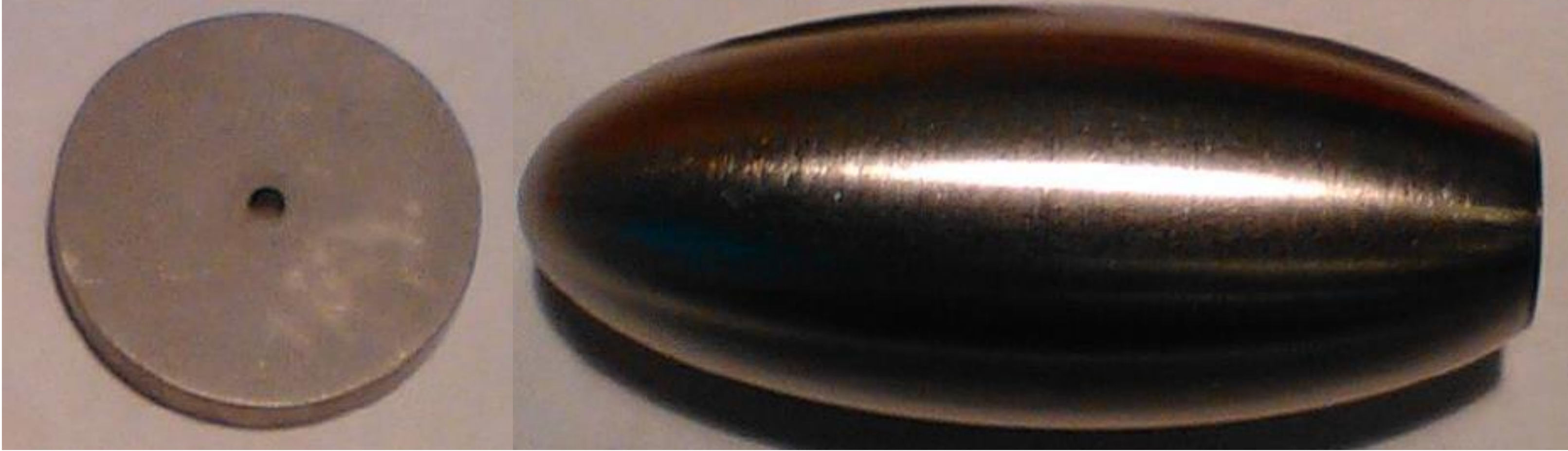}
	\caption{Left: Coin shaped sample used to measure the penetration depth in a LE-$\mu$SR experiment. Right: Ellipsoidal sample used to measure the DC field of first flux penetration with the surface $\mu$SR technique. Note the shiny surface of the Nb substrate before coating (right) while Nb$_3$Sn has a matt finish (left).}
	\label{fig:Samples}
\end{figure}

In the solid ellipsoid the Meissner state is supported by screening currents that augment the field at the equator and reduce the field at the poles. When the flux at the equator reaches the field of first vortex penetration $H\msub{vp}$, which can be $H\msub{c1}$ or in case of a surface barrier up to $H\msub{sh}$, fluxoids will nucleate at the equator and redistribute uniformly inside the superconductor due to vortex repulsion. Pin-free ellipsoidal bulk samples produce a uniform vortex flux density in the mixed state since the inward directed driving force on the vortex by the surface screening currents is compensated by the vortex line length that increases for fluxoids closer to the ellipsoid axis. The magnetic field at the equator  $H\msub{equator}$ will be enhanced to $H\msub{equator}=H\msub{a}/(1-N)$ where $H\msub{a}$ is the applied field and $N$ the demagnetizing factor. In our geometry the ellipsoid demagnetizing factor is $N=0.13$, therefore the measured field of first penetration $H\msub{a$\mid$entry}=0.87H\musb{vp}$. If the temperature is above \unit[9.25]{K}, the critical temperature of niobium, the geometry is a superconducting Nb$_3$Sn shell. The field will still break in at $H\msub{a$\mid$entry}=0.87H\musb{vp}$ but the vortex lines will directly snap to the center since the flux line length in the superconducting shell is actually less near the ellipsoid axis.
The coin sample is tested in the Meissner state using the LE-$\mu$SR technique. Due to its geometry the field is enhanced by \unit[15]{\%} at the edge and \unit[9]{\%} in the midplane. For details refer to \cite{Junginger_muSR_Overview}. The samples used for $\mu$SR are shown in \refF{fig:Samples}.

\subsection{QPR sample}
\begin{figure}[ht]
  \centering
  \includegraphics[width=0.90\columnwidth]{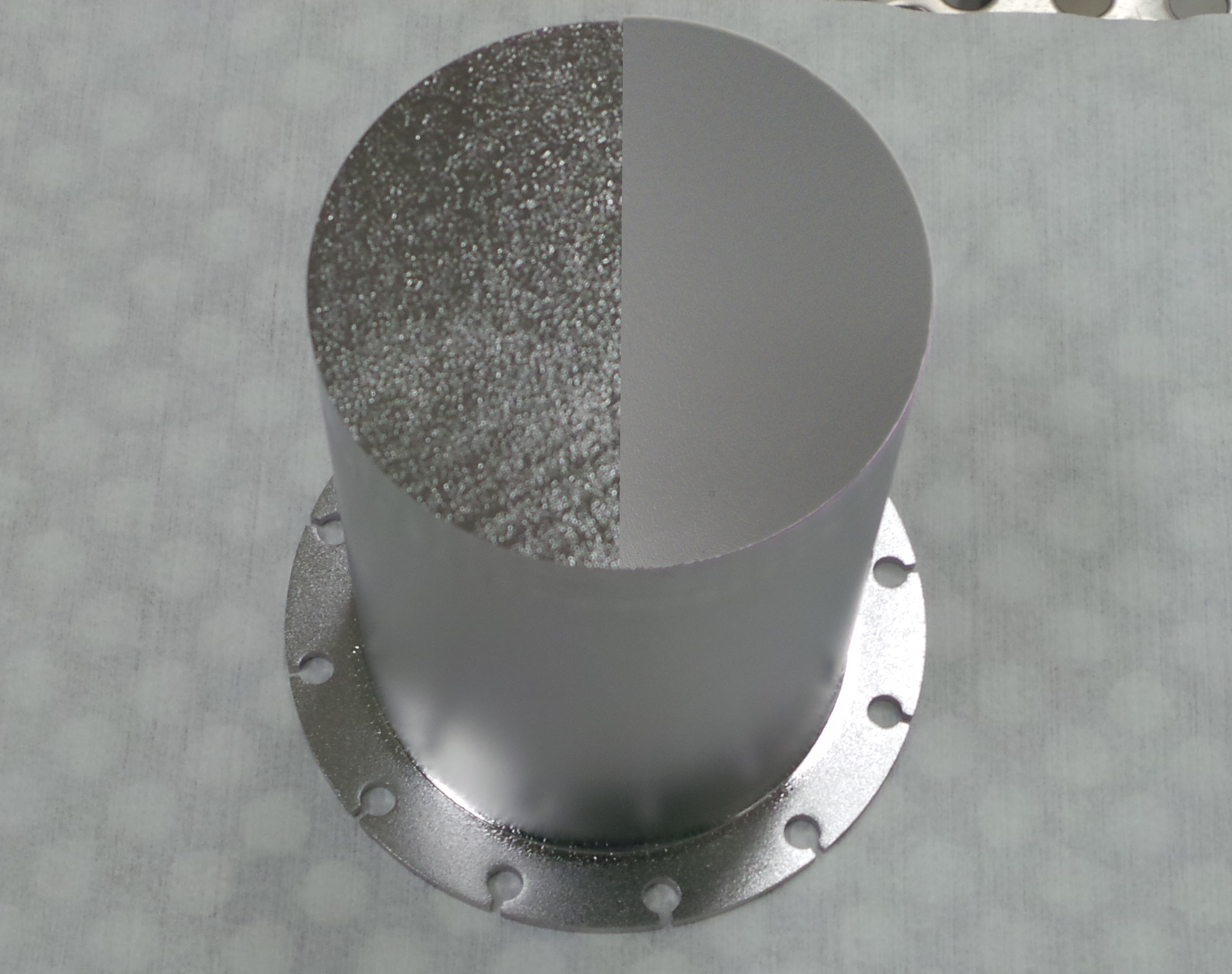}
  \caption{QPR sample used to measure the critical RF field. The stainless steel flange connected to the bottom end of the sample chamber is not shown. On the RF surface two pictures are stitched together, showing the Nb substrate before coating (left) and the Nb$_3$Sn coating (right).}
  \label{fig:Samples:QPR}
\end{figure}

The QPR sample chamber has a ``top hat''-like shape with its upper outside being exposed to the RF field inside the resonator, see \refF{fig:Samples:QPR}. The actual sample is a disk of \unit[75]{mm} diameter and \unit[10]{mm} in height. It is electron-beam welded to a tube of \unit[2]{mm} wall thickness which in turn is welded to the ``brim''. The bulk material of all parts is fine grain niobium with the Nb$_3$Sn coating covering all outer (and inner) surfaces. For mounting into the QPR the superconducting part of the sample chamber has to be connected to a double-sided CF100 flange. Commonly, QPR samples are brazed to this stainless steel flange. The process of coating with Nb$_3$Sn requires a substrate of high purity niobium, furthermore, heat related changes to the sample due to brazing or welding after coating have to be prevented. In order to enable coating with Nb$_3$Sn according to the process used for SRF cavities, a detachable connection of superconducting sample chamber and stainless steel bottom flange was implemented. An indium wire gasket of \unit[1]{mm} diameter was used providing both thermal link to the liquid helium bath and UHV compatible sealing of the cavity vacuum. Prior to coating with Nb$_3$Sn the niobium substrate was characterized at HZB showing low residual rf surface resistance of about $\unit[4]{n\ohm}$ and high RF critical field $\mu_0H\msub{c,RF}=\unit[220]{mT}$ \cite{phdkleindienst}. For details on the QPR refer to \cite{qprhzb2015,phdkeckert}.

\section{Methods}
\subsection{The $\mu$SR technique} 
In a $\mu$SR experiment, \unit[100]{\%} spin polarized muons are implanted one at a time into the sample, stop in interstitial sites and precess with characteristic Larmor frequencies determined by the local magnetic field $H\msub{int}$. When the muon decays it emits a fast positron which has a propensity to emerge along the direction of the muon spin. Two positron detectors (symmetrical placed around the sample) detect the positron signal. The time evolution of the muon spin polarization $P(t)$ gives information about the local field experienced by the muon. For a general introduction to $\mu$SR refer to \cite{yaouanc2011muon, Schenk}. If the field has fully penetrated, or the measurement is performed in the London layer using the LE-$\mu$SR technique, an oscillating depolarization will be observed 
\begin{equation}
P(t)=\exp{\left( -\frac{1}{2}\sigma^2t^2\right) }\cdot \cos{\left( \omega t + \frac{\pi\phi}{180}\right) }
\label{eq:Pol1}
\end{equation}
with 
\begin{equation}
\omega=\gamma_\mu H\msub{int},
\label{eq:omega}
\end{equation}
where $\gamma_\mu=2\pi$\unit[13.55]{kHz/G} is the gyromagnetic ratio of the muon and $\phi$ a phase which can depend on the external field. The leading term of \refE{eq:Pol1} describes the damping of the oscillation assuming a Gaussian distribution of the local field around $H\msub{int}$ with a full width at half maximum $\sigma$.

If the sample is in the Meissner state and the muon is implanted in the bulk with the surface muon beam, the muon does not sense the external magnetic field. The local magnetic field sensed in this case is from the nuclear dipolar fields. In case of significant muon diffusion, which is the case for high RRR niobium \cite{Junginger_muSR_Overview}, the polarization is described by the dynamic zero field Kubo-Tuyabe function $P_{\rm ZF}^{\rm dyn}$ \cite{Hayano} . Therefore, in the field of first vortex penetration measurements $P_{\rm ZF}^{\rm dyn}$ will be sensed for low applied field. When the field has fully penetrated the sample \refE{eq:Pol1} describes the depolarization. If the field penetrates gradually, which is the case if there is pinning, there can be coexisting field free and penetrated areas. In this case, the total polarization function is comprised of two terms as
\begin{eqnarray}
\label{eq:Asy}
P(t)&=&f_0\cdot P_{\rm ZF}^{\rm dyn}(t) + \\ \nonumber
& & f_1\cdot\exp{\left( -\frac{1}{2}\sigma^2t^2\right) }\cdot \cos{\left( \omega t + \frac{\pi\phi}{180}\right) }.
\end{eqnarray}

The first term is due to the volume fraction being in the Meissner state and the second term describes the volume fraction containing magnetic field. These fractions are represented by the coefficients $f_0$ and $f_1$, which can take values from 0 to 1. At zero magnetic field inside the sample $f_0$ equals 1 and $f_1$ equals 0. A significant decrease of $f_0$ from its low field value equal to 1 signifies vortex penetration. For details refer to \cite{Junginger_muSR_Overview}.
 
Two different muon beams are used for these studies. At TRIUMF so called surface muons (kinetic energy \unit[4.1]{MeV}) are used with an implantation depth of about \unit[130]{$\mu$m} deep in the bulk of a niobium sample. If measured above the critical temperature of niobium, the geometry of the samples can therefore be considered as a superconducting shell and the screening of the Nb$_3$Sn coating alone can be determined. This allows to detect the transition from the Meissner to a vortex state. For details refer to \cite{Junginger_muSR_Overview}.

At PSI, a low energy muon beam is available \cite{prokscha2008new,morenzoni2000low}. By varying its energy, it can be used as a probe for local magnetism at various depths between a few and up to about \unit[300]{nm} depending on the material density. Applying a field below $H\msub{c1}$ this allows to measure the local field $H\msub{int}$ as a function of depth. This experiment therefore allows for a direct measurement of the London penetration depth. 

\subsection{Quadrupole Resonator (QPR)}
The QPR \cite{Mahner:611593, Junginger:Revscientinstr12} is a dedicated sample test cavity which can be used to measure the RF quench field as a function of temperature. It is operated in a bath cryostat filled with superfluid helium which is pressure stabilized providing stable RF conditions. Regarding the sample chamber, only the stainless steel flange at its bottom is in contact with liquid helium. The RF sample disk is cooled via heat conduction through the sidewalls of the sample chamber. The bottom side of the sample disk is equipped with a resistive DC heater in its center and a calibrated Cernox temperature sensor at the radial position of highest RF field. Using the heater, arbitrary sample temperatures between 1.8 and approx. $\unit[25]{K}$ can be achieved. The heater power is PID controlled, providing temperature stability better than $\pm\unit[2]{mK}$ at \unit[18]{K}. Due to the weak thermal link to the liquid helium bath and the low thermal conductivity of the stainless steel flange, the maximum temperature difference along the RF surface is about \unit[0.1]{K} at \unit[18]{K}. Note that only the sample temperature is changed, whereas the helium bath temperature is kept at \unit[1.8]{K} throughout the entire measurement.

\begin{figure}[htb]
  \centering
  \includegraphics[width=0.95\columnwidth]{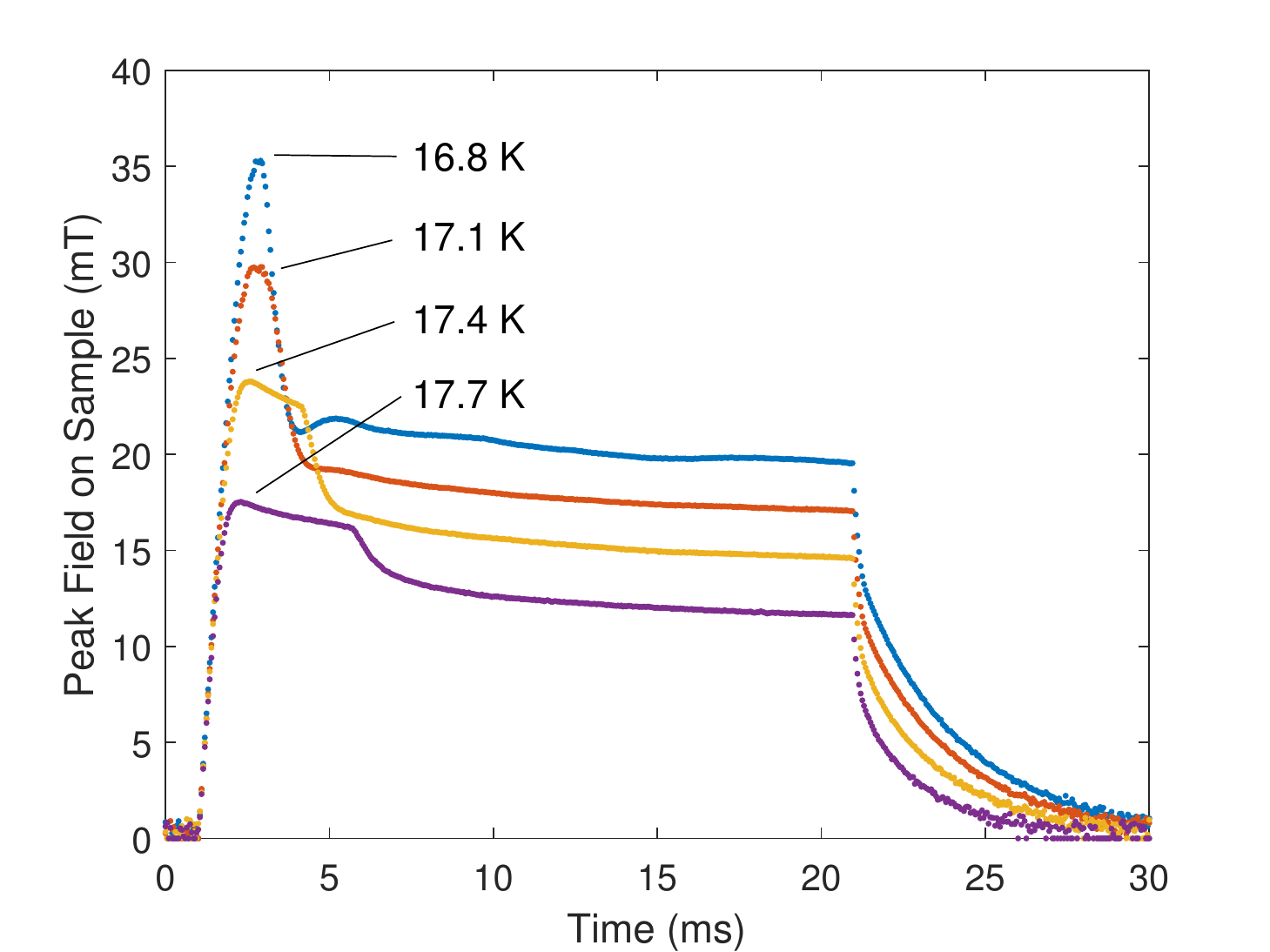}
  \caption{Instantaneous RF field on the sample surface measured by the pickup probe at four different temperatures. A constant pulse length of \unit[20]{ms} was used. The RF quench field is given by the peak value. The post-quench behavior changes with temperature since it depends on temperature dependent thermal properties of the sample and on RF characteristics like Lorentz-force detuning and the loaded quality factor of the QPR.}
  \label{fig:qpr-trace}
\end{figure}

In order to probe the RF critical field of the sample, single rectangular RF pulses are used. A quench decreases the loaded quality factor of the resonator, leading to a drop of transmitted RF power. The transmitted power is recorded with a time resolution of $\unit[50]{\mu s}$, allowing for a precise determination of the peak RF field. \refF{fig:qpr-trace} shows four examples of transmitted RF pulse traces, converted to peak RF field on the sample. The temperature measurement is not fast enough to capture the temporal evolution of the sample but it is used to differentiate between a quench occurring on the sample or somewhere else inside the resonator.

The field configuration inside the QPR provides an advantage over tests using elliptical cavities: In case of a \unit[1.3]{GHz} TESLA-shaped single-cell cavity the ratio of peak magnetic field and square root of stored energy is given by $\frac{B\msub{pk}}{\sqrt{U}}=\unit[36.5]{\frac{mT}{\sqrt{J}}}$. Hence, a stored energy of $U=\unit[7.5]{J}$ is required to achieve $B\msub{pk}=\unit[100]{mT}$. For the QPR this is reduced significantly:
\begin{equation}
  U\msub{QPR} = \left(\frac{\unit[100]{mT}}{\unitfrac[319]{mT}{\sqrt{J}}}\right)^2 = \unit[0.098]{J}
\end{equation}
In order to suppress pre-quench heating the RF time constant $\tau=Q\msub{L}/\omega$ has to be small, but reducing the loaded quality factor $Q\msub{L}$ increases the required RF drive power. Assuming the cavity to be strongly overcoupled, the required forward power is given by $P\msub{f}=\frac{U}{4\tau}$. This yields for the case of equal rise time and magnetic field:
\begin{equation}
  \frac{P\msub{f, TESLA}}{P\msub{f, QPR}}=\frac{U\msub{TESLA}}{U\msub{QPR}}=\frac{7.5}{0.098}=76
\end{equation}
Clearly, nearly two orders of magnitude more RF power must be applied to the TESLA cavity to achieve the quench field.
The maximum achievable field level inside the QPR is limited by the cavity quench field at about \unit[120]{mT}. Hence, the critical field can only be measured at elevated temperatures with $R\msub{s}$ being dominated by BCS losses. Using $P\msub{diss}\propto R\msub{s}\propto\omega^2$ yields for localized heating
\begin{equation}
  \frac{P\msub{diss, TESLA}}{P\msub{diss, QPR}} = \frac{\omega\msub{TESLA}^2}{\omega\msub{QPR}^2} = \left(\frac{\unit[1.3]{GHz}}{\unit[414]{MHz}}\right)^2 \approx 10.
\end{equation}
For this reason, the requirement of equal rise time in QPR and cavity measurements can be relaxed which in turn reduces the necessary RF power. Furthermore, time-resolved numerical simulations of the QPR sample chamber show a significant impact of the sample thickness on the temperature dynamics \cite{phdkeckert}. Due to the thermal diffusivity of niobium, the sample disk being \unit[10]{mm} in thickness reduces the non-equilibrium temperature rise during the RF rise time compared to cavity-grade material of $2$-\unit[3]{mm} thickness. Note that the scenario of a thin sample disk is approximately comparable to the situation of a single-cell cavity test at temperatures above \unit[4.2]{K} with inefficient cooling by gaseous helium only. Hence, at HZB a solid-state amplifier with a few hundred watts is sufficient, while measurements with TESLA cavities had to be performed with MW-class klystrons \cite{phdkeckert,hays1997,posen2015prl}. 

\section{Results}
\subsection{Low energy muon spin rotation}
\begin{figure}[t]
  \centering
    \includegraphics[width=0.9\columnwidth]{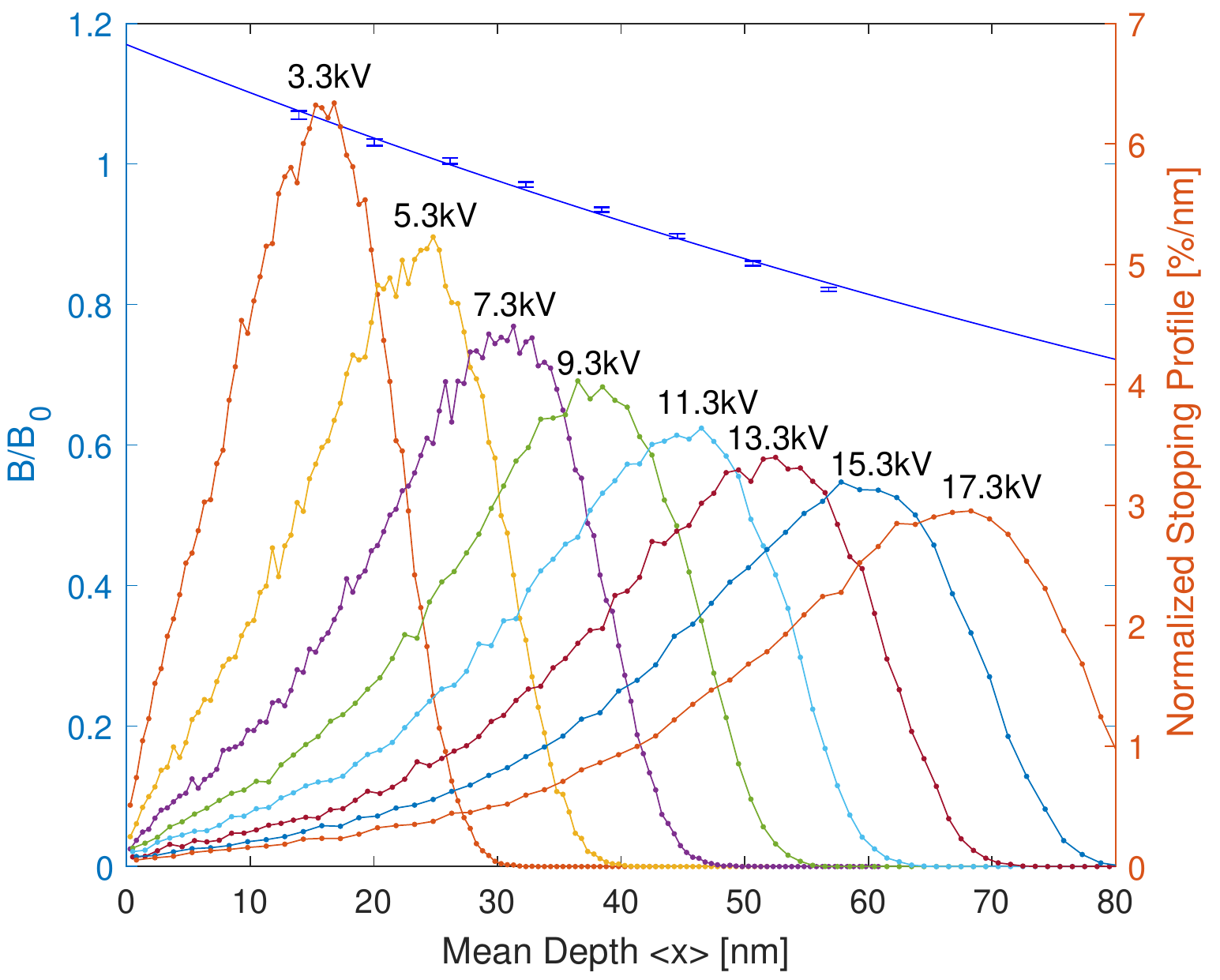}
    \caption{Left y-axis: Normalized local magnetic field as a function of muon implantation depth at $T$=\unit[5]{K} measured with low energy muon spin rotation. Right y-axis: Normalized muon stopping profile as a function of depth $\avg{x}$ obtained from the Monte-Carlo code \texttt{TRIM.SP}. This program has been developed at MPI Garching by W.\,Eckstein \cite{biersack1984sputtering, ecksteincomputer} and adopted and experimentally verified for muons \cite{morenzoni2002implantation}.}
  \label{fig:BvsE}
\end{figure}

Muons have been implanted with energies between 3.3 and \unit[17.3]{keV} corresponding to mean stopping depths between 15 and \unit[55]{nm}, see \refF{fig:BvsE}. The applied field was $\mu_0 H_a$=\unit[10]{mT} and the temperature \unit[5]{K}. For each depth the local magnetic field sensed by the muons is derived from the time evolution of the muon spin polarization function P(t) using \refE{eq:Pol1}.
The data has been analyzed using the musrfit software \cite{Suter}. The whole data was also collectively fitted to the one-dimensional London model using the BMW libraries for musrfit \cite{BMW_libs} taking into account the stopping distribution of the muons in the material. For an isotropic local superconductor a solution of the form   
\begin{equation}
B=B_0 \exp{\left(-x/\lambda\right)}
\end{equation}
is obtained, where $x$ is the distance from the surface. The fit parameters found are \ $B_0$=\unit[11.4]{mT} and $\lambda(5K)$=\unit[160.5(3.9)]{nm}. $\lambda$ is expected to change according to
\begin{equation}
\lambda(T)=\frac{\lambda(0K)}{\sqrt{1-{\left(T/T_c\right)}^4}},
\label{eq:GorterCasimir}
\end{equation}
from which $\lambda(0K)$=\unit[160.0(3.9)]{nm} can be derived being only \unit[0.3]{\%} shorter than $\lambda(5K)$. The lower critical field $H\msub{c1}$ can be expressed as \cite{tinkham2004introduction}
\begin{equation}
\mu_0H\msub{c1}=\frac{\Phi_0}{4\pi\lambda^2}\ln(\kappa+0.5),
\label{eq:Hc1}
\end{equation}
where $\Phi_0$ is the magnetic flux quantum. To obtain the Ginzburg-Landau parameter $\kappa=\lambda/\xi_{GL}$ from the measured value of $\lambda$, literature values of the London penetration depth $\lambda_L$ and the BCS coherence length $\xi_0$ are required. A formula is derived using the fact that the BCS and the Ginzburg-Landau coherence length are both correlated to the magnetic flux quantum $\Phi_0$ \cite{tinkham2004introduction} 
\begin{equation}	
 \kappa=\frac{\lambda(T,l)}{\xi\msub{GL}}=\frac{2\sqrt3}{\pi}\frac{\lambda^2}{\xi_0\lambda\msub{L}}.
	\label{eq:kappatf}
\end{equation}
Note that $\xi_0$ and $\lambda_L$ are both fundamental material properties defined for the clean stoichiometric material. The values taken from literature can be found in \refT{tab:MaterialParameters}. From $\lambda(0K)$=\unit[160]{nm}, $\kappa$=60(15) and $\mu_0 H\msub{c1}$=\unit[28(2)]{mT} are derived. The thermodynamic critical field can be calculated using \cite{tinkham2004introduction}
\begin{equation}
H\musb{c}=\frac{\sqrt{2}\kappa H\msub{c1}}{\ln{\kappa}}.
\end{equation}
The value of \unit[600(100)]{mT} is consistent with \unit[520]{mT} reported in \cite{Godeke2006}, see \refT{tab:MaterialParameters}. Finally, the superheating field is calculated using \refE{eq:Hsh} to be \unit[500(120)]{mT}. On a side note the value of $B_0$ is \unit[14]{\%} larger than the applied field. This field enhancement is close to what is expected at the edge of this coin shaped sample ($N$=0.15)\footnote{A more thorough analysis would have to take the complete magnetic field and muon stopping distributions into account.}.

\subsection{Surface muon spin rotation}

\begin{figure}[b]
 \centering
	 \includegraphics[width=0.9\columnwidth]{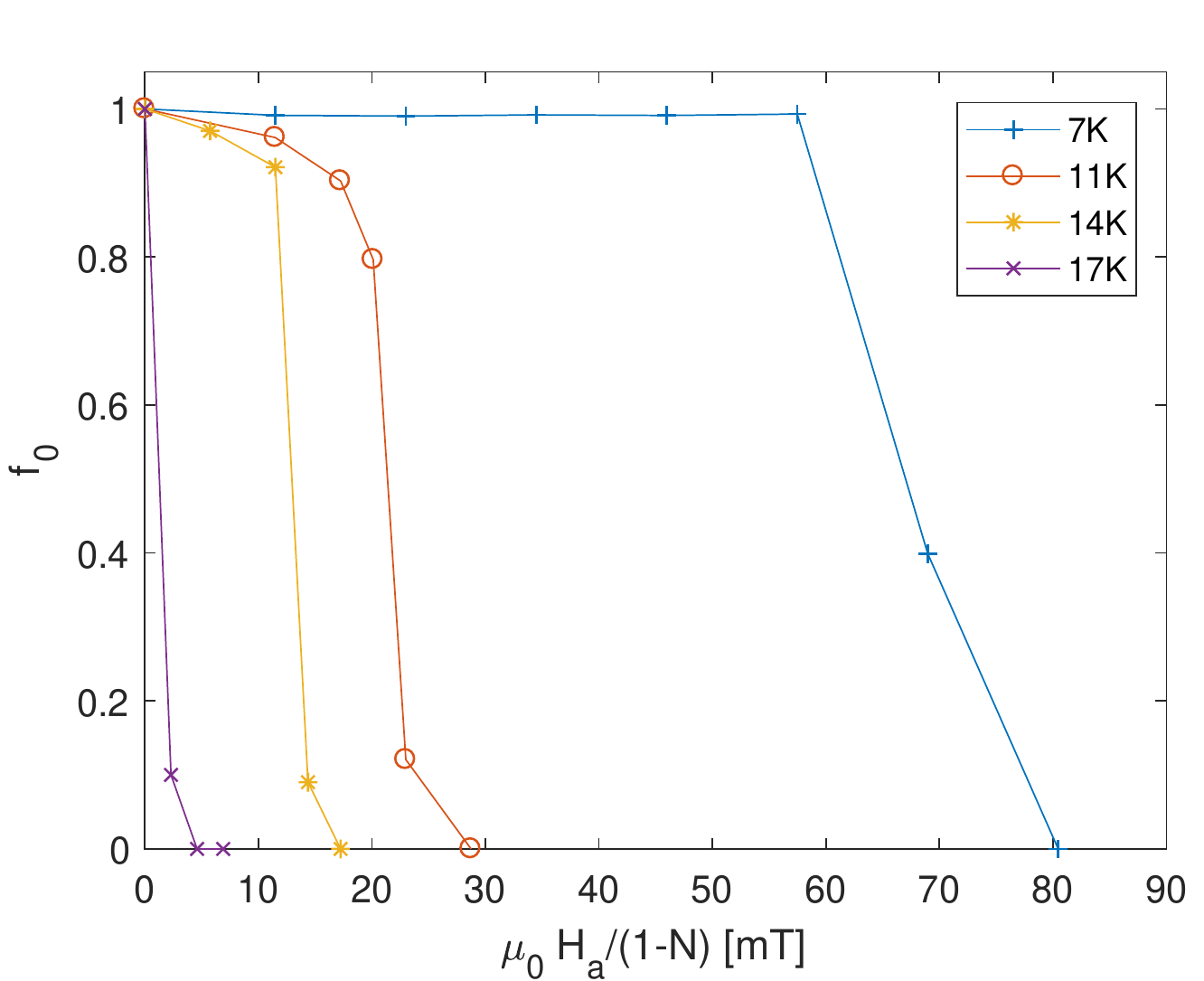}
   \caption{Fit parameter $f_0$ signifying the volume fraction probed by the muons which is in the field free Meissner state as a function of applied field at different temperatures. Below \unit[9.2]{K} the niobium bulk material provides additional shielding yielding a strong increase of the field of first vortex penetration.}
	\label{fig:fvsB}
\end{figure}

The muon polarization function was measured as a function of applied field at three temperatures above and one below $T\msub{c}[Nb]$ using surface muons implanted $\approx$\unit[130]{$\mu$m} in the bulk. \refF{fig:fvsB} displays the fit parameter $f_0$ as derived from \refE{eq:Asy} as a function of applied field divided by 1-$N$, where $N$ is the demagnetization factor of the ellipsoid. From this plot the intrinsic DC field of first vortex penetration $H\musb{vp,DC}$ can be derived for each temperature as the field for which $f_0$ significantly deviates from 1. For temperatures above $T\msub{c}[Nb]$ the geometry is a \unit[2]{$\mu$m} superconducting shell, while below $T\msub{c}[Nb]$ both materials can potentially provide shielding and it is not possible to directly derive the field of first vortex penetration for the Nb$_3$Sn layer alone. The field of first vortex penetration for the Nb$_3$Sn shell at \unit[0]{K} is derived by extrapolation taking only data above \unit[9.25]{K} into account. A value of $\mu H\msub{vp}$=\unit[28(12)]{mT} is found, see discussion.

\vfill

\begin{figure}[htb]
  \centering
  \includegraphics[width=\columnwidth]{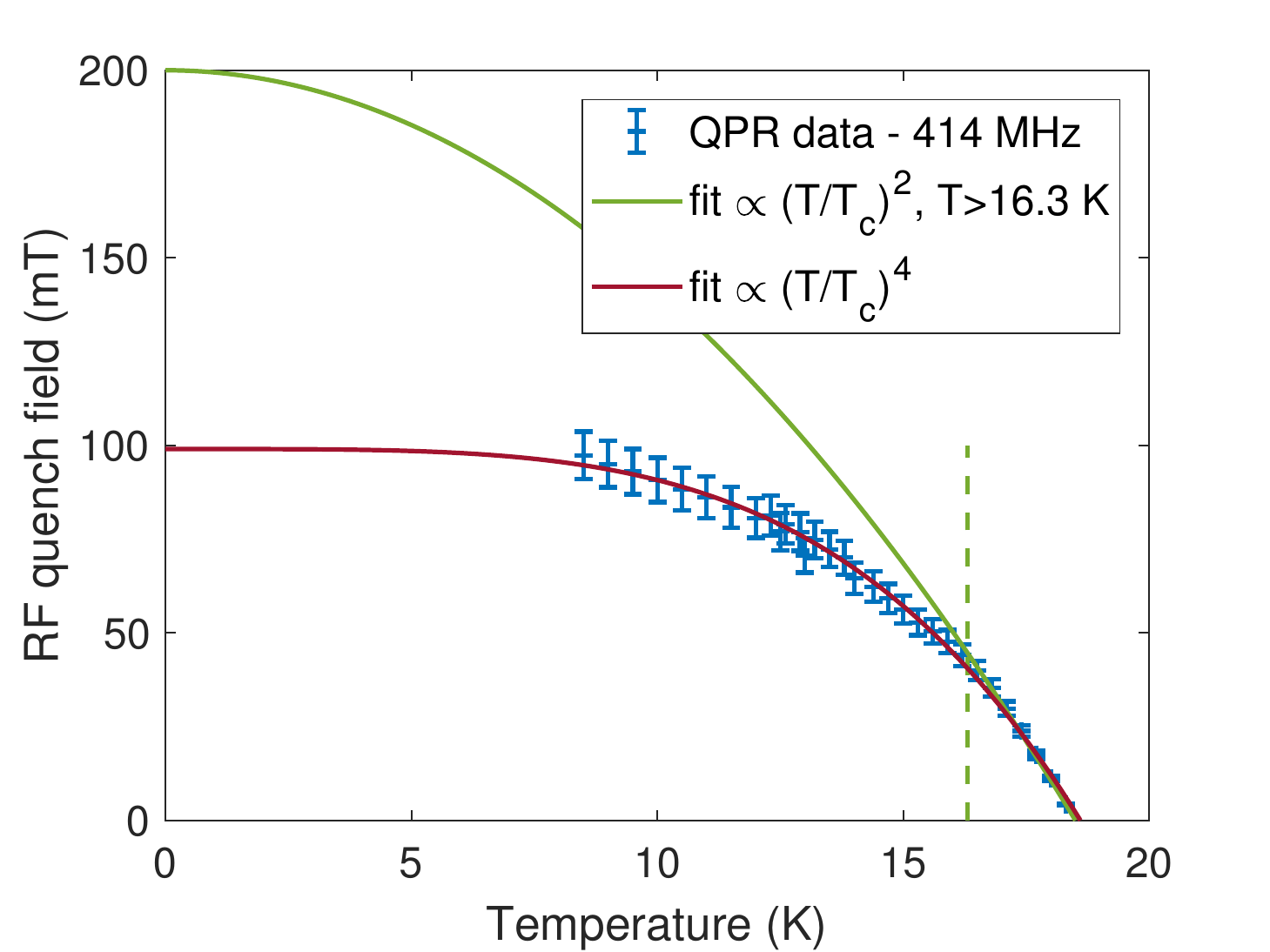}
  \caption{RF quench field measurement data. The full data range can be described by a temperature dependence $\propto (T/T\msub{c})^4$. For the fit according to \refE{eq:HcRF} only data with $T\!>\!\unit[16.3]{K}$ is used, as highlighted by a dashed line.}
  \label{fig:qprhcrit}
\end{figure}

\begin{figure}[h]
  \centering
  \includegraphics[width=\columnwidth]{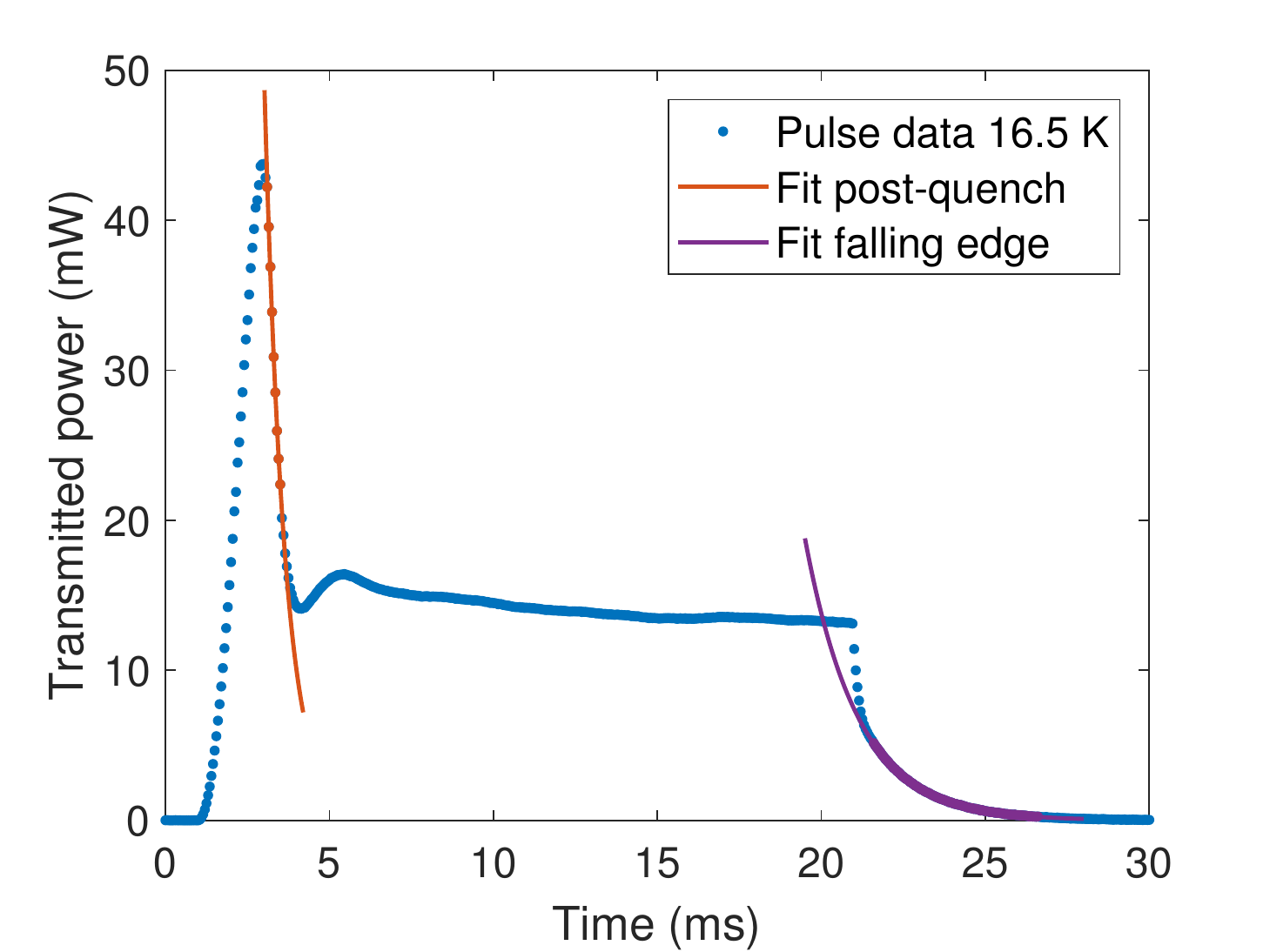}
  \caption{Typical pulse trace with fits for quenched decay time $\tau\msub{quench}$ (orange decay curve) and unperturbed falling edge decay time $\tau\msub{L}$ (purple decay curve).}
  \label{fig:qpr-pulsefit}
\end{figure}

\newpage

\subsection{Quadrupole Resonator}
The RF quench field was measured at various temperatures $T\!>\!\unit[8]{K}$, see \refF{fig:qprhcrit}. This temperature limitation was due to the fact that RF instabilities and the quenching resonator ruled out measurements for $B\msub{pk}\!>\!\unit[100]{mT}$.

\begin{figure}[b]
  \centering
  \includegraphics[width=\columnwidth]{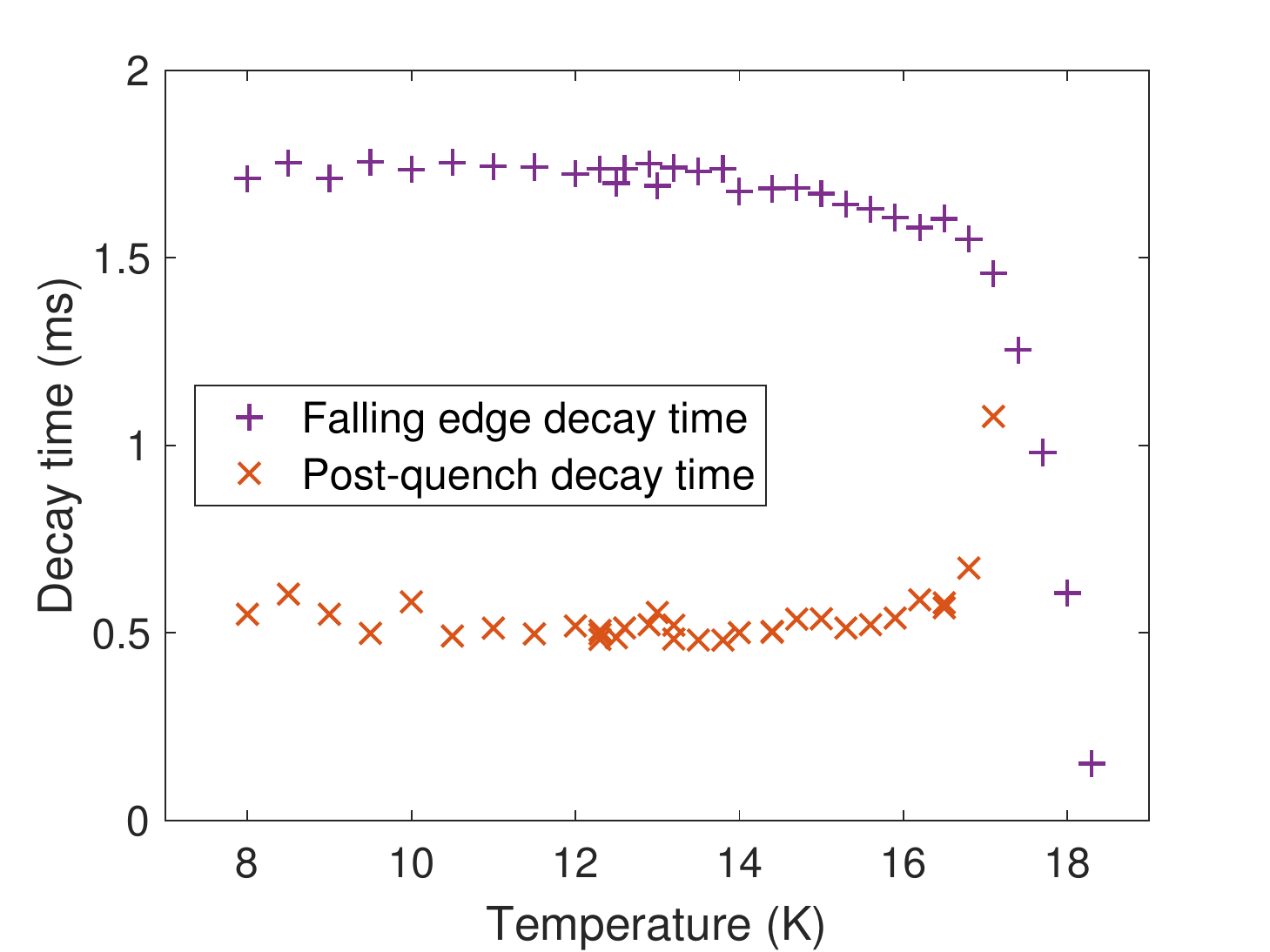}
  \caption{Unperturbed and post-quench decay times as a function of sample temperature. Note that for $T\!>\!\unit[16]{K}$ the increasing surface resistance of the sample reduces $\tau\msub{L}$. For $T\!>\!\unit[17]{K}$ no post-quench decay time can be derived.}
  \label{fig:qpr-decaytime}
\end{figure}

The measured data can be described by the empirical relation for the temperature dependence of the critical thermodynamic field 
\begin{equation} \label{eq:HcRF}
H\msub{c}(T) = H\msub{c}(0)\cdot\left(1-\left(\frac{T}{T\msub{c}}\right)^2\right)
\end{equation}
only in the temperature range $T\!>\!\unit[16.3]{K}$. This choice is somewhat arbitrary, but expanding the fit data range to lower temperature reduces the fit quality significantly. Extrapolation to low temperature yields $\mu_0H\msub{vp,RF}$=$\unit[200(5)]{mT}$ and $T\msub{c}$=$\unit[18.5(1)]{K}$. Note that the predicted temperature dependence for $H\msub{sh}$ is almost identical to that of $H\msub{c}$.

In order to estimate a possible impact of pre-quench RF heating, which would falsify the assumed sample temperature, the acquired pulse shapes are investigated. After a quench, the loaded quality factor of the QPR is dominated by the partially normal conducting sample. The corresponding decay time $\tau\msub{quench}$ is determined by fitting the drop of transmitted power (see \refF{fig:qpr-pulsefit}). After the RF drive power is switched off, the sample becomes superconducting within few $\unit[100]{\mu s}$ and the following unperturbed decay time $\tau\msub{L}$ is again obtained from an exponential fit (see \refF{fig:qpr-pulsefit}).

From $\tau\msub{quench}$ and the electrical resistivity of Nb$_3$Sn in the normal conducting state ($\unit[8-20]{\mu\Omega\,cm}$ \cite{valente2016}), the quenched sample area is estimated yielding $\unit[2-3]{mm^2}$. Note that for $T\!>\!\unit[16]{K}$ the increasing temperature dependent surface resistance of the sample begins to reduce the decay time $\tau\msub{L}$ of the superconducting state (see \refF{fig:qpr-decaytime}). For $T\!>\!\unit[17]{K}$, the change of loaded quality factor decreases such, that no steep decrease of transmitted power is observed anymore and fits of $\tau\msub{quench}$ are meaningless to deduce the quenched sample area. The increase observed for the \unit[17]{K} point in \refF{fig:qpr-decaytime} hence is considered to reflect this measurement difficulty. This is also visible in the pulse traces shown in \refF{fig:qpr-trace}.

However, the fact that for $T\!<\!\unit[17]{K}$ $\tau\msub{quench}$, and hence the quenched area, do not depend on temperature (see \refF{fig:qpr-decaytime}) is a strong indication of a localized quench spot without excessive pre-quench heating affecting the measurement accuracy. This is in agreement with a quench limitation set by vortex line nucleation (see the following discussion). In comparison, measurements of a bulk niobium sample with RF quench fields consistent with the global superheating limit showed a quench area of about $\unit[100]{mm^2}$ \cite{phdkeckert}.

\section{Discussion}
In this study the vortex penetration fields of Nb$_3$Sn for both DC and RF magnetic fields have been measured. In addition, we derived the theoretical limits $H\msub{c1}$ and $H\msub{sh}$. 

The lower critical field $H\msub{c1}$ is the field at which it is energetically favorable for flux to be located inside the superconductor instead of being expelled by Meissner currents. An energy barrier can prevent flux penetration up to the superheating field $H\msub{sh}$. If it is unknown whether or not there is a surface barrier, neither of these two fields can be directly measured. Here $H\msub{c1}$ and $H\msub{sh}$ are both obtained from the magnetic penetration depth as derived by low energy muon spin rotation. The results listed in \refT{tab:Results} are in good agreement with data derived from low field surface impedance measurements \cite{posen2015prl}. $H\msub{c1}$ is found to be consistent with the DC field of first vortex penetration measured with surface muons.

\begin{table}[b]
  \centering
  \caption{Vortex penetration and critical fields of Nb$_3$Sn extrapolated to \unit[0]{K} in mT divided by $\mu_0$. $H\msub{vp,RF}$ and $H\msub{vp,DC}$ are obtained from extrapolation to \unit[0]{K}. For $H\msub{vp,DC}$ and the first value of $H\msub{vp,RF}$ a quadratic temperature dependence was assumed as in \refE{eq:HcRF}. For the latter only temperatures above \unit[16.3]{K} have been taken into consideration. The second value of $H\msub{vp,RF}$ is extrapolated from a fit to \refE{eq:vlnm} using the full data set. All other values are obtained from material parameters with the RF cavity values taken from \cite{posen2015prl} and \cite{phdposen}.} \label{tab:Results}
  \begin{tabular}{l|c|c||c|c|c}
  Method    & $H\msub{vp,RF}$ & $H\msub{vp,DC}$ & $H\msub{c1}$ & $H\msub{sh}$ & $H\msub{c}$ \\ \hline
  $\mu$SR   &        -        &    28(12)       &    28(2)     &   500(120)   &  620(120)   \\
  QPR       &    200(5)       &      -          &      -       &      -       &     -       \\
            &    100(3)       &                 &              &              &             \\
  RF cavity & $\approx250$    &      -          &    29(2)     &   400(80)    &  600(100)   \\
  \end{tabular} 
\end{table}

\begin{figure}[h]
  \centering
  \includegraphics[width=\columnwidth]{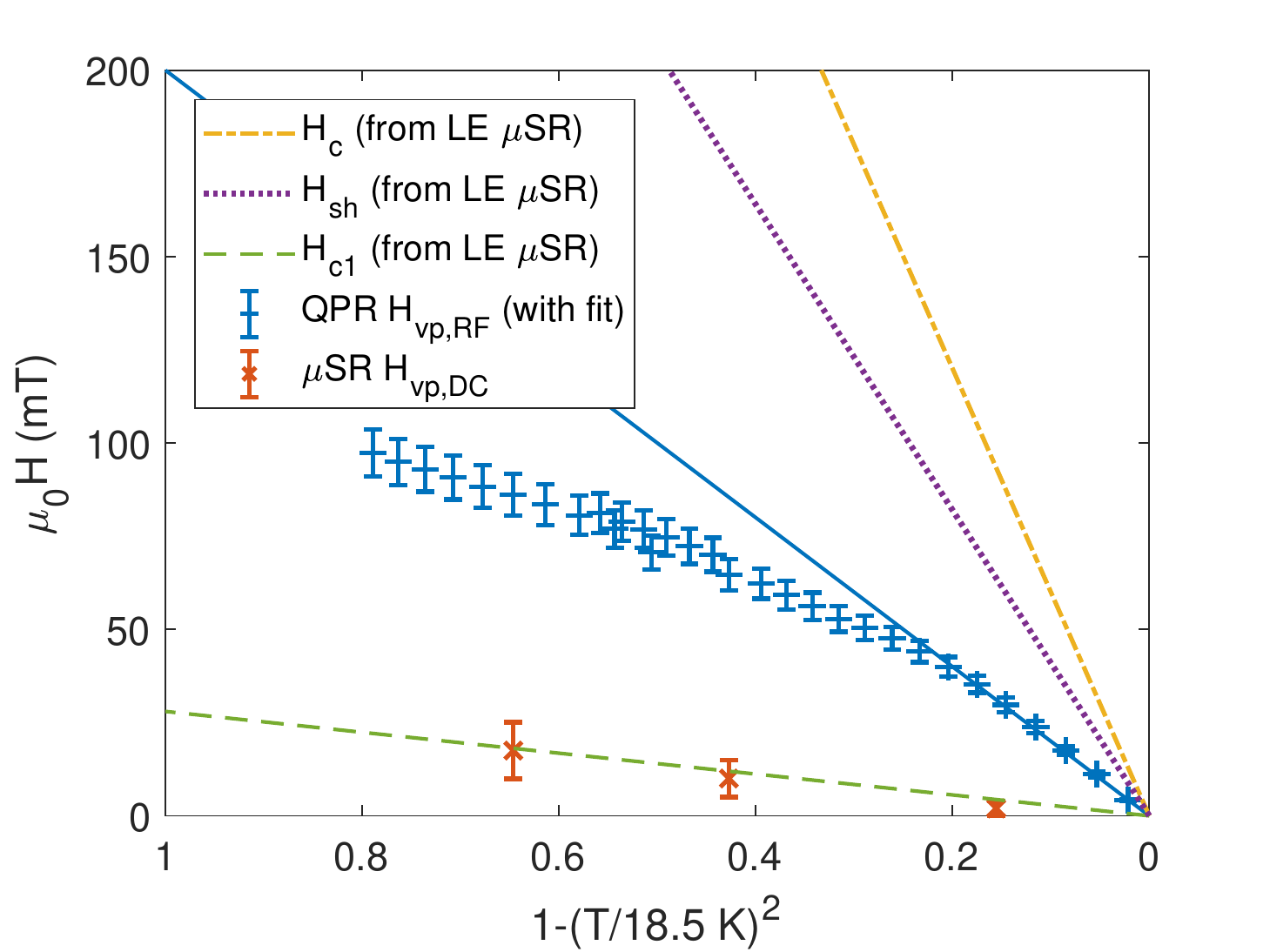}
  \caption{Combined plot showing data for $H\msub{c}$, $H\msub{sh}$ and $H\msub{c1}$ derived from low energy $\mu$SR measurement assuming a temperature dependence according to \refE{eq:HcRF}, and field of first vortex penetration as obtained from QPR (RF) and surface-$\mu$SR (DC) measurements. For the QPR data the quadratic fit from \refF{fig:qprhcrit} is shown again.}
  \label{fig:CriticalField}
\end{figure}

The RF field of first vortex penetration $H\msub{vp,RF}$ has been measured as a function of temperature at \unit[414]{MHz} using a Quadrupole Resonator. The values of $H\msub{vp,RF}(T)$ clearly exceed $H\msub{c1}$ and $H\msub{vp,DC}(T)$, see \refF{fig:CriticalField}. This signifies that a metastable state above $H\msub{c1}$ has been reached. However, the extrapolated value of $\mu_0H\msub{vp,RF}(\unit[0]{K})$=$\unit[200(5)]{mT}$ is about a factor of two below the predicted superheating field of this material as listed in \refT{tab:MaterialParameters}. As a thermal limitation is unlikely, one may argue that the superheating field is locally suppressed, potentially by coating flaws as reported in \cite{trenikhina2017performance}. Another cause of reduced vortex penetration field might be surface roughness. Recent analysis of Nb$_3$Sn surfaces showed roughness causing local field enhancement exceeding \unit[50]{\%} \cite{porter2016linac}. In that case, the measured $H\msub{vp,RF}$ should not reach the theoretical values, but be suppressed by a factor of about 1/1.5.

A possible explanation for measurement data $H\msub{vp,RF}(T)$ deviating from the expected quadratic behaviour is given by vortex penetration at defects with a size of the order of the coherence length. The vortex line nucleation model (VLN) \cite{Yogi,saito2003srf} gives an heuristic formula for this limitation
\begin{equation} \label{eq:vlnm}
  H\msub{VLN}(T)=H\msub{VLN}(0)\cdot\left(1-\left(\frac{T}{T\msub{c}}\right)^4 \right).
\end{equation}
The temperature dependence is consistent with this model for the entire range (see \refF{fig:qprhcrit}). Restricting the fit range to the same high-temperature range as used for the fit to \refE{eq:HcRF} (see \refF{fig:qprhcrit}) still yields a higher adjusted $R^2$-value of $R^2$=$0.9997$ for the fit to \refE{eq:vlnm}, compared to $R^2$=$0.9979$ for the quadratic model from \refE{eq:HcRF}.

\newpage

\section{Conclusion}

In this study the maximum RF and DC fields of Nb$_3$Sn prepared for SRF application have been measured using a Quadrupole Resonator and the muon spin rotation technique. These results have been compared to theoretical field limitations, i.e. the lower critical field $H\msub{c1}$ and the superheating field $H\msub{sh}$. As both quantities depend on the London penetration depth $\lambda$, this material parameter has been directly measured using the low energy muon spin rotation (LE-$\mu$SR) technique. The different techniques required samples of different shape and size but all samples have been prepared in the same coating chamber under same conditions.

The combined experimental results strongly confirm that Nb$_3$Sn SRF cavities can indeed be operated in a flux free Meissner state above $H\musb{c1}$. It is suggested that localized vortex penetration with little or negligible preheating prevents current Nb$_3$Sn SRF cavities to reach higher field values. This might be interpreted as a local suppression of the superheating field potentially at coating flaws as reported in \cite{trenikhina2017performance}.

While in cavity tests $\lambda$ is often extracted from low field surface impedance measurements, here LE-$\mu$SR was used for the first direct measurement of this parameter on Nb$_3$Sn prepared for SRF application. This allows to give a better determination of $H\msub{c1}$ and $H\msub{sh}$ than what has previously been reported. Furthermore, the strong agreement of the low energy muon spin rotation results with cavity data from \cite{posen2015prl} gives confidence in the latter method of extracting material parameters from low field surface impedance measurements for future studies with different coating parameters potentially resulting in a different $\lambda$.

\section{Acknowledgements}
This work has been funded partly by a Marie Curie International Outgoing Fellowship of the European Community's 7th Programme. The QPR part of this project has received funding from the European Union's Horizon 2020 Research and Innovation programme under Grant Agreement No 730871. The Cornell activities of this project were supported by the U.S. Department of Energy under Award {DE-SC0008431} and by the U.S. National Science Foundation under Award {PHY-1549132}, the Center for Bright Beams. The LE-$\mu$SR measurements were performed at the Swiss Muon Source (S$\mu$S), at the Paul Scherrer Institute in Villigen, Switzerland. Thanks to Gerald Morris, Bassam Hitti, Donald Arseneau, Deepak Vyas, Rahim Abasalti and Iain McKenzie from the TRIUMF CMMS support team.


\end{document}